# Direct Observation of Second Order Atom Tunnelling

S. Fölling[1], S. Trotzky[1], P. Cheinet[1], M. Feld[1], R. Saers[2], A. Widera[1,3], T. Müller[1,4] and I. Bloch[1]

[1]*Institut für Physik, Johannes Gutenberg-Universität, 55099 Mainz, Germany*

[2]*Department of Physics, Umeå University, 90187 Umeå, Sweden*

[3]*Institut für Angewandte Physik, Universität Bonn, 53115 Bonn, Germany*

[4]*Institute of Quantum Electronics, ETH Zürich, 8093 Zürich, Switzerland*

**Tunnelling of material particles through a classically impenetrable barrier constitutes one of the hallmark effects of quantum physics. When interactions between the particles compete with their mobility through a tunnel junction, intriguing novel dynamical behaviour can arise where particles do not tunnel independently. In single-electron or Bloch transistors, for example, the tunnelling of an electron or Cooper pair can be enabled or suppressed by the presence of a second charge carrier due to Coulomb blockade[1,2]. Here we report on the first direct and time-resolved observation of correlated tunnelling of two interacting atoms through a barrier in a double well potential. We show that for weak interactions between the atoms and dominating tunnel coupling, individual atoms can tunnel independently, similar to the case in a normal Josephson junction. With strong repulsive interactions present, two atoms located on one side of the barrier cannot separate[3], but are observed to tunnel together as a pair in a second order co-tunnelling process. By recording both the atom position and phase coherence over time, we fully characterize the tunnelling process for a single atom as well as the correlated dynamics of a pair of atoms for weak and strong interactions. In addition, we identify a conditional tunnelling regime, where a single atom can only tunnel in the presence of a second particle, acting as a single atom switch. Our work constitutes the first direct observation of second order tunnelling events with ultracold atoms, which are the dominating dynamical effect in the strongly interacting regime. Similar second-order processes form the basis of superexchange interactions**

between atoms on neighbouring lattice sites of a periodic potential, a central component of proposals for realizing quantum magnetism[4-7].

For the description and observation of quantum mechanical tunnelling, a double-well type potential, where two localized spatial modes are separated by a barrier, is among the conceptually simplest setups. When a particle is initially prepared on one side of this barrier, it will tunnel back and forth between the two sides with a well-defined frequency. For macroscopic quantum systems such as superconductors or atomic Bose-Einstein condensates this tunnel coupling can lead to a Josephson type tunnelling dynamics[8-10]. When interactions between individual particles are much stronger than the tunnel coupling in the system, quantized Josephson dynamics arises, where e.g. in superconducting devices the charge carriers tunnel individually across barriers[11,12]. In the case of coupled mesoscopic quantum dots, a co-tunnelling regime can be achieved, where separate electrons only tunnel in a correlated way[13,14]. For ensembles of ultracold atoms in periodic potentials, strong interactions fundamentally alter the properties of the many body system, leading to strongly correlated phases such as the Mott insulating state[15-19]. In such cases, where direct first order tunnelling of single atoms is highly suppressed, second-order correlated tunnelling processes can be the dominant dynamical effects. Despite the absence of direct long-range interaction mechanisms between particles, second-order "superexchange" type processes can provide effective spin-dependent interactions between particles on separate positions[4-7].

The dynamics of interacting bosonic atoms in a double well with tight confinement is described by a quantized Josephson or two-mode Bose-Hubbard Hamiltonian[11,12]

$$H = -J\left(\hat{a}_L^\dagger \hat{a}_R + \hat{a}_R^\dagger \hat{a}_L\right) - \frac{1}{2}\Delta\left(\hat{n}_L - \hat{n}_R\right) + \frac{1}{2}U\left(\hat{n}_L(\hat{n}_L - 1) + \hat{n}_R(\hat{n}_R - 1)\right), \quad (1)$$

with $J$ the tunnelling matrix element, $\hat{a}_{L,R}^\dagger$ and $\hat{a}_{L,R}$ the creation and annihilation operators for a bosonic particle in the ground state of the left or right well, $\Delta$ the bias potential between the wells, and $U$ the interaction energy for two particles in a single well. In the following we assume repulsive interactions ($U>0$); however, attractive interactions lead to a similar evolution. The operators $\hat{n}_L$ and $\hat{n}_R$ count the number of atoms in the left and right mode, respectively. In the tight confinement approximation, the quantum state of the atoms can be described in a Fock state basis $|n_L, n_R\rangle$ with $n_L$, $n_R$ being non-negative integers. For a single atom, the two possible states $|1,0\rangle$ and $|0,1\rangle$ are coupled by the tunnel matrix element $J$. For the case of two atoms, the states $|2,0\rangle$ and $|0,2\rangle$ both directly couple to the state $|1,1\rangle$ via the tunnelling term in first order. For strong interactions $U \gg J$, the energy difference between the states $|2,0\rangle$ and $|1,1\rangle$ is much larger than the coupling, resulting in a strong detuning and therefore suppressed transitions between these states. For an unbiased junction ($\Delta=0$), the state $|0,2\rangle$, however, always has the same energy as $|2,0\rangle$. The direct transition to this state, which corresponds to the co-tunnelling of both atoms as a pair, is therefore still resonant. This second order tunnelling process has an effective matrix element $2J^2/U$, which can be obtained by second order perturbation theory for $J/U \ll 1$ (see refs. [5,6]).

In order to realize the double well potentials for ultracold rubidium atoms, we superimpose two periodic potentials with a periodicity of 382.5 nm (short lattice) and 765.0 nm (long lattice) and controllable intensities and relative phase (see methods). They are produced by two independent optical standing waves in such a way that each potential minimum of the long lattice is split into two wells by a maximum of the short lattice potential (see Fig. 1a). By changing the relative phase of the two potentials, a controlled bias $\Delta$ can be introduced, which can be approximated by the energy difference between the two potential minima. Additional standing waves on the two orthogonal axes provide transverse confinement, creating a three-dimensional array of up to $10^5$ double wells[20] occupied by one to two $^{87}$Rb atoms each. In this array we carry out many identical instances of the experiment in parallel to obtain the quantum mechanical expectation values of the observables in a single run.

The initial state with the atoms localized on one side of the double wells is obtained by adiabatic changes of the potential after loading in the symmetric configuration. Both sites are combined and the atoms are brought to rest on the left side of their respective double well (see Fig. 1b). The tunnelling dynamics can subsequently be initiated by quickly reducing the barrier height within 200 μs. After an evolution period, we determine the resulting average position and the average single particle phase relation between the quantum states on the left and right well. The overall sums of atoms $N_L$ and $N_R$ in the left and the right wells, respectively, are obtained by rapidly suppressing tunnelling again in 200 μs and employing a band mapping technique[21] (see Fig. 1c). We calibrate this method by making reference measurements with fully localized atoms prepared on either side. From the two occupation numbers the average position $\langle x \rangle = (N_R - N_L)/(N_R + N_L)$ of the particles is calculated, which denotes their position relative to the barrier in units of $d/2$, where $d$ is the well separation. The phase relation between the two wells is determined by a separate interferometric sequence, where the lattice is switched off rapidly at the end of the dynamical evolution. The emerging double-slit matter wave interference pattern then yields the average single particle phase relation (see methods and Fig. 1d).

Since approximately 40% of the population is in singly-occupied double wells, we determine the atom pair signal in a two-stage process. First the data point which includes both singly and doubly-occupied wells ("total signal") is recorded for a given configuration. The measurement is then repeated with a "filtering sequence" before the dynamical evolution period, which removes atom pairs from the trap (see methods). By this, we obtain the single-atom signal, which can then be subtracted from the total signal to obtain the data point for atom pairs.

The time evolution for atom pairs can be calculated using the Hamiltonian (1) for any given set of double well parameters $U$, $J$ and $\Delta$ within a three-state. Our model for the ensemble takes into account inhomogeneities in the parameter $\Delta$ as well as in the atom density due to the overall confining potential (see methods). We observe experimentally that the parameters in the Bose-Hubbard description in the case of pair occupation are slightly, but notably modified. Specifically, the effective tunnel coupling has to be described by a modified effective single-particle tunnelling rate $J'$ for the two-atom Hamiltonian, which is 3-10% higher than the free-particle $J$. This is in agreement with estimates using perturbative modifications to the wave function caused by the interactions.

The measured time-resolved traces resulting from single atom and atom pair tunnelling are shown in Figure 2. The single atom datasets (black dots) show the expected sinusoidal population oscillation between the two wells at a frequency $2J/h$. If the interaction energy $U$ is smaller than the tunnel matrix element $J$ ($J/U$~1.5, Fig. 2a and b), tunnelling of a single atom out of a pair is only slightly detuned. This process therefore competes with the resonant second order tunnel process, leading to a signal containing more than one frequency component (see Fig. 2b). When reducing $J$ to reach the interaction-dominated regime ($J/U$~0.2, Fig. 2c-f), the single-atom signal is still sinusoidal at a correspondingly lower frequency. However, the average position of atom pairs now shows a strongly modified behaviour. The pair-breaking first-order process is highly suppressed due to the detuning by the interaction energy $U$, and is visible as a small modulation with a period of ~400 μs (Fig. 2d). In contrast, the second-order hopping process is now the dominant dynamical effect, leading to a much slower oscillation with a period of ~1.8 ms. Tunnelling is resonant in the centre of the trap, however, in double wells located away from the centre along the double well axis, the overall confining potential leads to a bias $\Delta \neq 0$. This detuning is especially significant in the case of the second-order process, where the potential bias enters twice and the effective tunnel coupling is low. Thus, for low coupling strengths, the observed oscillation amplitude of the slow second-order process averaged over the ensemble is lower than 1, but remains the dominant dynamical process.

In addition to the average centre of mass position, the phase and visibility as obtained from the interferometric sequence are shown (Fig. 2 e,f). The average direction and velocity of the flow of atoms are characterized by the phase and visibility observables and the tunnelling parameter $J$. For the single atom case, one observes distinct jumps in the phase by π, whenever the particle is localized to one side of the potential well and reverses its propagation direction. When atom pairs are included, a much more complex phase evolution emerges. By fitting the modelled dynamical evolution both for atom pairs and single atoms to the data, the tunnelling matrix elements and the interaction energies can be extracted. Figure 3 shows the fitted tunnelling frequency $2J/h$ against the short lattice depth $V_{short}$ in units of the recoil energy $E_r = h^2/2m\lambda^2$, with $m$ being the mass of the atoms, $h$ Planck's constant and $\lambda$=765 nm the short lattice wavelength. We fit both $J$ and $J'$ independently and find good qualitative agreement with the theoretical prediction for $J$. However, we typically measure a 5% lower coupling than predicted by a band structure calculation, a deviation which is slightly larger than expected from the uncertainties of the lattice depths. The resulting pair tunnelling frequency $(4J'^2/U)/h$ is also plotted, showing the much faster decrease with growing barrier height compared to single particle tunnelling. For comparison, we plot the interaction strength $U$ as well as the expected frequency of the detuned first order tunnelling process $\sqrt{4J'^2+U^2}/h$, which asymptotically approaches $U$ as $J\rightarrow 0$.

The interaction-induced suppression of first order tunnelling for atom pairs is a consequence of energy conservation. Tunnelling can be made resonant again by biasing the single-atom ground state of the well into which the atom tunnels. If the energy offset $\Delta$ is equal to $U$, the first-order process is resonant in the presence of another atom, but detuned by $\Delta$ without a second atom. The signal for this conditional tunnelling process is depicted in the inset of Figure 4, where occupation by two atoms results in a significant amplitude of the sinusoidal tunnelling signal, whereas in the case of a single atom the dynamics is barely visible. The fitted amplitudes of both processes are plotted in Figure 4 against the bias energy, showing only one resonance at $\Delta$=0 for single atom occupation. The two-atom case has two resonances, one for the first order process at $\Delta$=$U$ and the other at $\Delta$=0, where the second order hopping process is in resonance.

In conclusion, we have reported on the first full characterization of the tunnelling dynamics of an interacting atom pair across a quantum weak link. By tuning the system from a weakly to a strongly interacting regime, we have been able to observe the transition from independent single particle tunnelling towards correlated tunnelling in second order exchange-type processes. Our measurement shows that for higher occupations per lattice site, corrections apply to the coupling parameters of the Hubbard model as determined for a single atom. Finally, an additional correlated tunnelling regime was demonstrated, in which a particle can tunnel only on the condition of a second atom being present. Such a single atom switch can be used to efficiently create entanglement over different lattice sites. Starting from spin-triplet pairs in single wells, created e.g. via spin-changing collisions[22], a resonant tunnelling event can create long-lived entangled singlet or triplet states[23]. Superexchange interactions between neighbouring spin pairs could then be used to engineer large correlated spin-chains, similar to those encountered in cluster states[24] or resonating valence bond-type states[25,26] of condensed matter physics.

**Methods**

**Double well lattice potential.** The light for the two standing waves along the double well axis is created by a 1530 nm fibre laser and a Titanium-Sapphire laser running at 765 nm. The frequency relation is controlled by frequency-doubling a small part of the 1530 nm output and recording the beat note between this light and the 765 nm lattice light. This beat note is stabilized to any offset frequency within a 1 GHz interval by a feedback circuit which adjusts the frequency of the fibre laser using a piezo-mechanical actuator. In our setup both laser beams are retro reflected by a common mirror, such that the relative phase of the two periodic potentials is fixed at the mirror position. The phase slip over the distance between mirror and atoms (25 cm) caused by choosing a non-zero offset frequency allows us to control the relative phase of the two standing waves at the position of the atoms, thereby choosing the well geometry and bias energy. Typical potential depths are 9.5 $E_r$ for the long lattice and 33 $E_r$ for each 843 nm transversal lattice beam, in units of the 765 nm recoil energy $E_r$, resulting in a parabolic confinement of ~ $2\pi \times 80$ Hz in the longitudinal direction.

**State preparation.** After creating an almost pure BEC of typically $8 \times 10^4$ atoms, we ramp up the optical potential within 160 ms. During the ramp, inter-double-well tunnelling is suppressed, but not the tunnelling through the barrier (10 $E_r$ final short lattice depth). This results in a majority of the atoms being loaded into doubly-occupied sites. We then adiabatically remove the central barrier to bring the atom pairs together in one well (Fig. 1b). This well is moved to the left by ½ of the short lattice period before the short lattice is ramped up again. By subsequently moving the long wavelength lattice back to the original position we arrive again at a symmetric potential with all atoms now on the left side of their respective double well.

**Filtering sequence.** In order to obtain the tunnelling signal for single atoms, we transfer all atoms from the $F=1$, $m_F=-1$ hyperfine state to the $F=2$, $m_F=0$ hyperfine state by a microwave adiabatic rapid passage after merging the wells. This strongly enhances spin relaxation collisions[22,27,28], which release enough energy to efficiently remove both atoms of each pair from the lattice. After a 40 ms hold time, the remaining atoms in singly occupied sites are transferred back to the $F=1$, $m_F=-1$ state.

**Model.** In order to predict the time evolution of an individual double well system, we diagonalise the Hamiltonian (1) for the single atom and the two-atom case independently and calculate the time evolution for the two initial states $|1,0\rangle$ and $|2,0\rangle$. Due to the Gaussian shape of all lattice beams, the tunnelling, interaction and bias parameters are not perfectly identical for all double wells of the array. We assume a shell structure distribution[29] of the atoms, with an outer region of singly occupied double wells and a spherical inner core of doubly occupied wells. This corresponds to a zero temperature assumption. For our chosen lattice ramps and double well configuration we do not expect a perfect shell structure, but the obtained dynamical evolution is not very sensitive to changes of the distribution shape. For example, using a thermal distribution of the atoms with temperatures comparable to $U/k_B$ and $J=0$ gives similar results. Here, $k_B$ denotes the Boltzmann constant. We only take into account the inhomogeneity of $\Delta$ across the cloud due to the harmonic confinement. The resulting dephasing of the ensemble is the only cause of damping of the total signal within our model. Other effects such as the inhomogeneity of the tunnelling matrix element as well as all inhomogeneities due to imperfections of the beam shapes are not included. We assume a fixed outer radius of 30 sites and use the trap frequency of the confinement and the radius of the inner shell as fit parameters. When fitting the model to the phase and visibility data from measurements, additional sources of damping can be present. These cause an overestimation of the trapping frequency obtained from the fit, which we observe to be up to 200%. The effect is most prominent towards the large $J/U$ regime, where the simple three-mode approximation starts to fail. The observed damping of the position signal (Fig. 2 a, b) in this regime cannot be reproduced with our model and is included by an additional, empirically determined exponential damping term with a decay time of 3.5 ms.

**Phase measurement and fitting.** In order to determine the phase relation between the wells, the trap is switched off and the interference pattern from the wave functions emerging from the two wells is recorded. The image is integrated perpendicularly to the double well axis and a double-slit interference pattern with Gaussian envelope is fitted to the resulting profile:

$$P(x) = A e^{-(x-x_0)^2/W^2} \cdot (1 + V\cos(kx + \theta)) \qquad (2)$$

The fit parameters of the envelope are the amplitude $A$, the width $W$ and average centre $x_0$. The interference pattern parameters are the visibility $V$, the periodicity $k$ as well as the phase θ, which is directly given by the average single-particle phase relation between the wells. The momentum-space Wannier-function envelope of the measured profiles is not perfectly Gaussian, especially for large tunnel coupling, leading to an overall decrease in visibility. For fitting the tunnelling dynamics model, the time-dependent signal obtained in this way is used. We simultaneously fit the signal for the total (mixed) population and for the filtered population of singly-occupied double wells as shown in Fig. 2. We do not perform the atom-number sensitive separation of the interference patterns into the signals from singly- and doubly occupied sites and model the mixed signal directly for fitting. Apart from the traces shown in Fig. 2, we typically do not record the average position for fitting, since the interference analysis yields more information per single image than the determination of the average position and the resulting data allows the determination of all parameters shown in Fig. 3.

We acknowledge helpful discussions with A. M. Rey and B. Paredes as well as funding through the DFG and the European Union (MC-EXT QUASICOMBS). R. S. acknowledges support by the EU QUDEDIS program as well as SJCKMS, Kempe I and II foundations.

Correspondence and requests for materials should be addressed to I.B. (e-mail: bloch@uni-mainz.de).

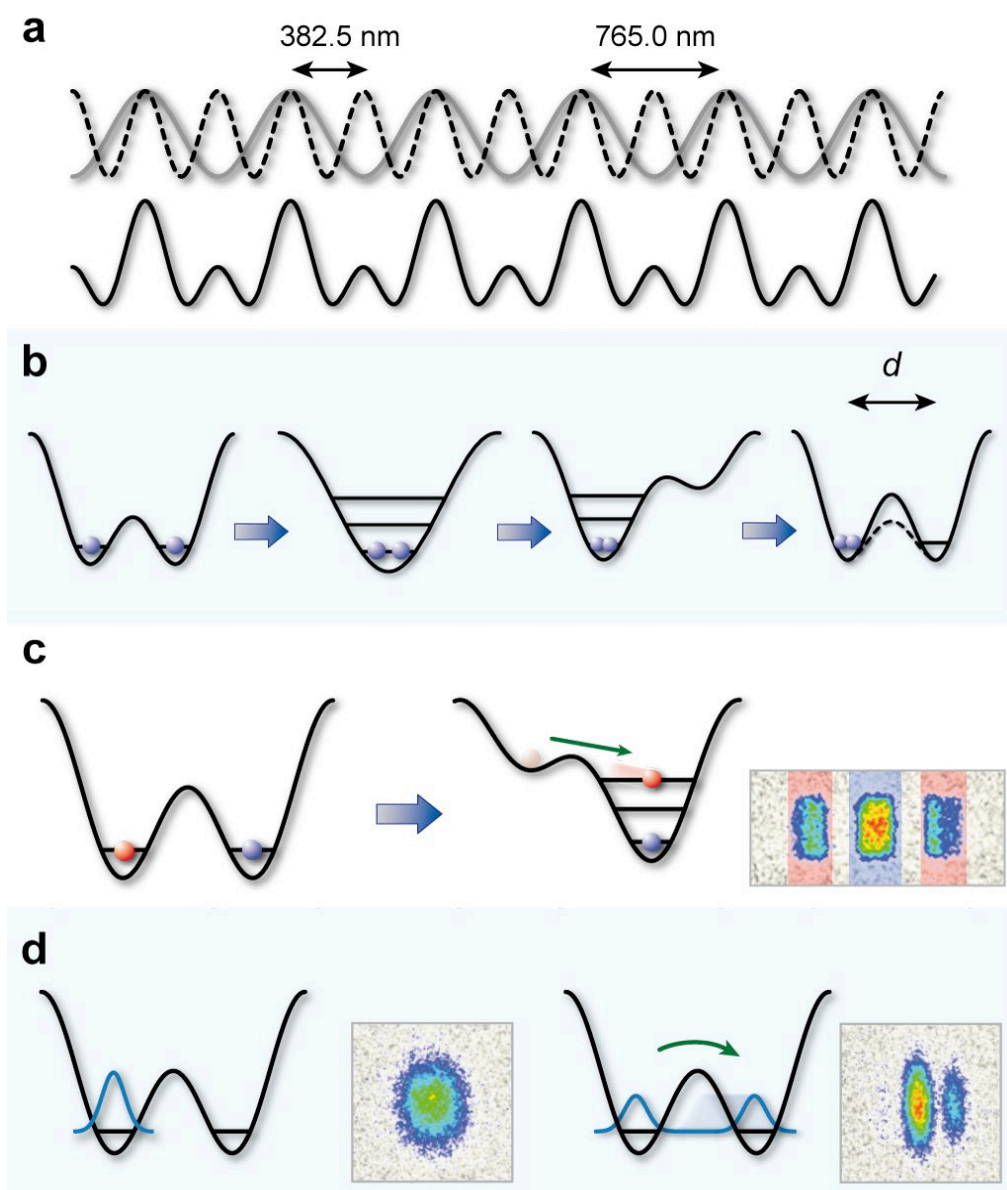

Fig. 1 Schematics of double well generation, loading and detection sequences. **(a)** Superimposing two optical lattice potentials differing in period by a factor of two creates an array of double well potentials. **(b)** Preparation sequence: An initially large well is split into a biased double well potential such that each left well is populated. The bias is then removed and the central barrier lowered to initiate the tunnelling dynamics (*d* denotes the well separation). **(c)** Position measurement. The atom number on each side can be recorded by 'dumping' the population of the left well into an excited vibrational state of the right well[21]. Subsequent band-mapping projects both states into separate Brillouin zones in free space[30] (marked red and blue in the figure inset). **(d)** Interferometric detection. After sudden release from the double well

potential and a period of free expansion, the double slit interference pattern is recorded. Particles localized to one well exhibit no interference; for delocalized atoms the pattern yields the relative single particle phase ($-\pi/2$ in the case shown).

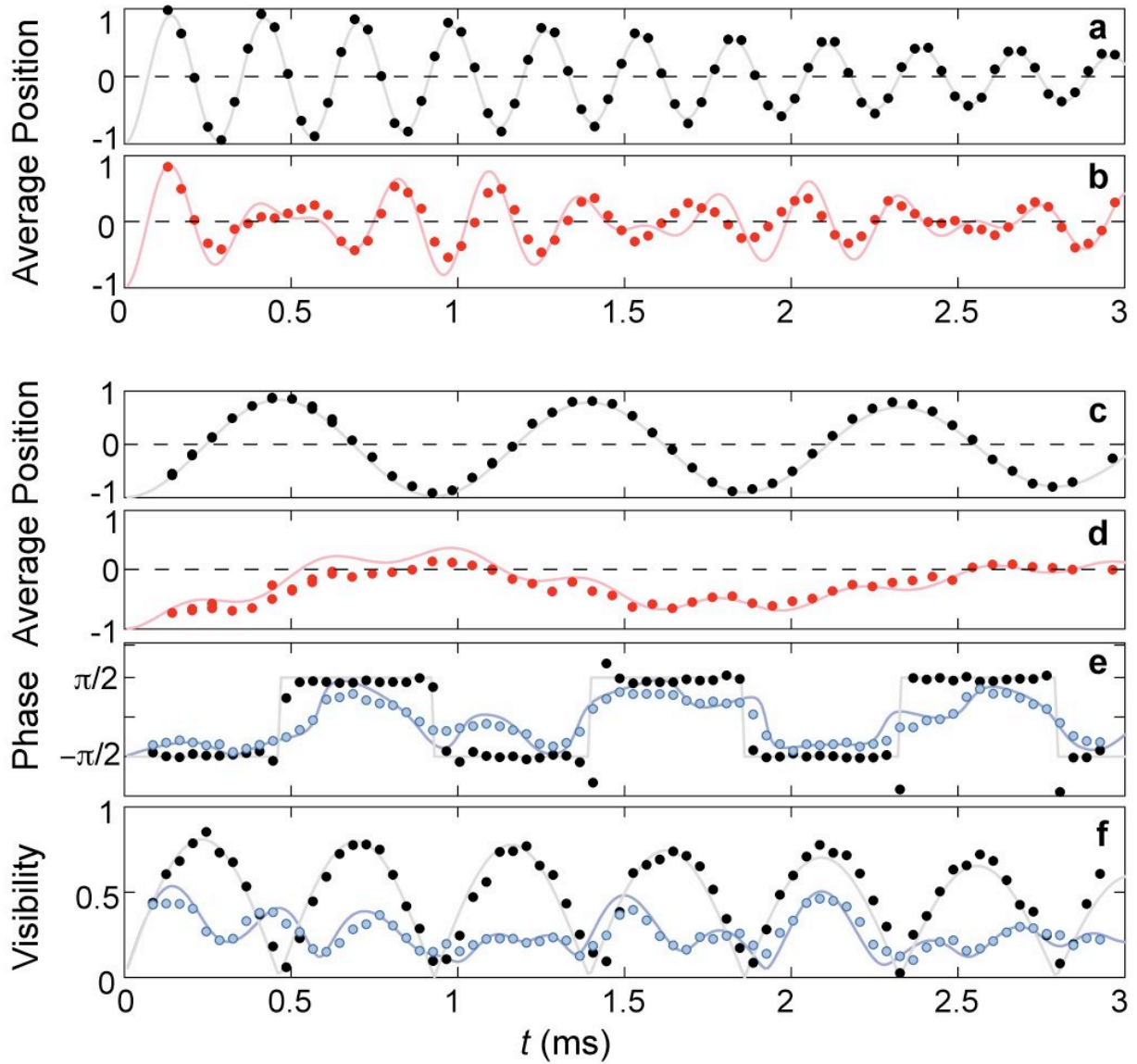

Fig. 2 Tunnelling dynamics. Full dynamical tunnelling evolution for single atoms and atom pairs in the weakly (*J/U*=1.5) **(a,b)** and strongly interacting regime (*J/U*=0.2) **(c-f)** after initially preparing all particles localized on the left side (position -1) of the double well. The black dots denote the single atom position **(a,c)**, phase **(e)** and visibility **(f)** signal. The red dots show the atom pair signal **(b,d)**. For strong interactions **(d)**, first-order tunnelling is suppressed and shifted to the detuned frequency 2.5kHz ~*U/h*. The main dynamical process is pair tunnelling with frequency $4J^2/hU$~550Hz  Blue hollow dots in **(e,f)** denote the combined single atom and atom pair signal recorded in the experiment by the interferometric detection method. The solid lines are fits to the data using a model based on the quantized Josephson Hamiltonian (see text).

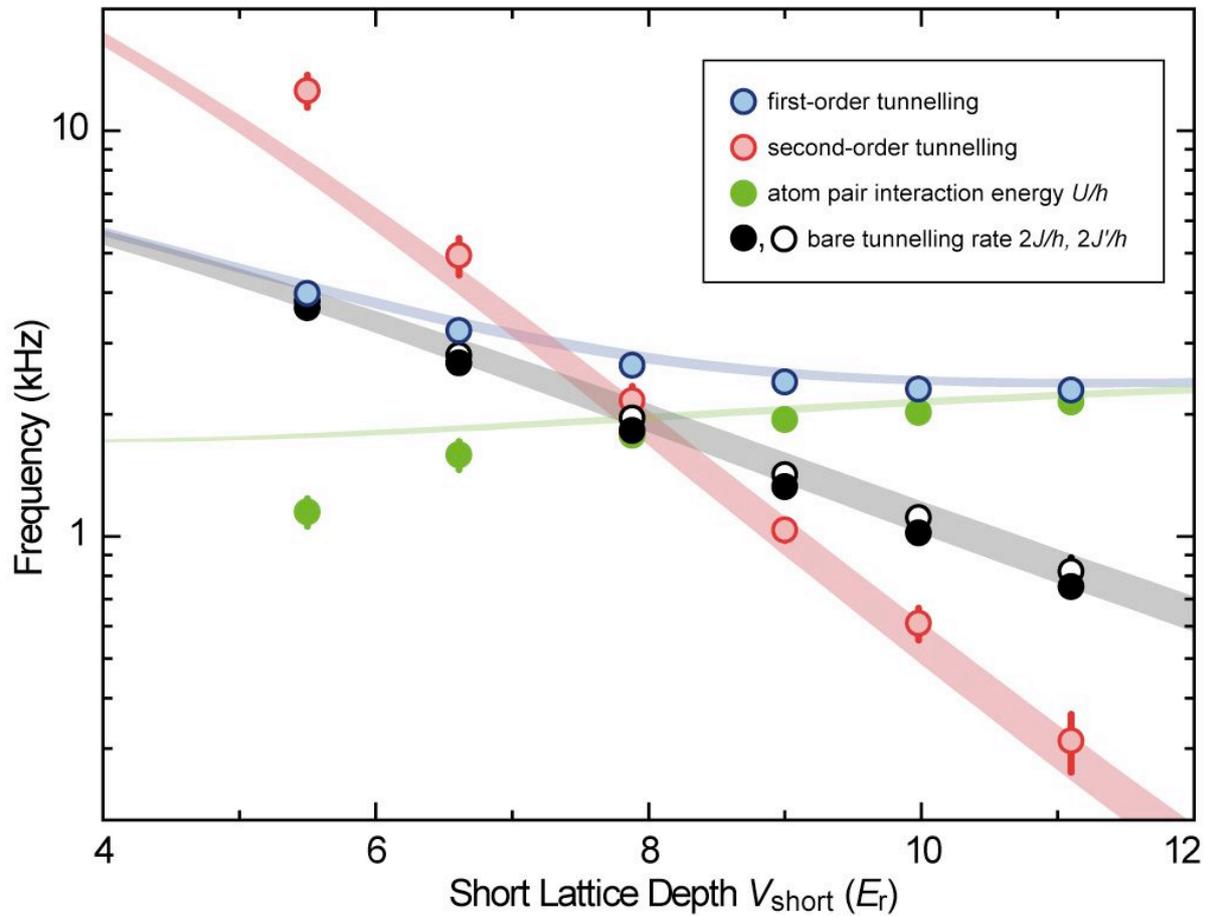

Fig. 3 Tunnelling frequencies vs. short lattice depth (barrier height). The frequencies corresponding to the coupling matrix elements $J$ and $J'$ (see text) are denoted with black dots and circles, respectively. The crossover above which the measured interaction energy $U$ (green filled circles) dominates over kinetic energy takes place around 8 $E_r$. The characteristic frequencies for the second order tunnelling process and the first order tunnelling for atom pairs derived from these values are shown as red and blue hollow dots, respectively. Error bars denote the 90% confidence intervals as determined from the fits. The shaded areas show the calculated frequencies for the single atom tunnelling as well as for the first and second order tunnel process as determined from band structure calculations. Their width represents the systematic 2% uncertainty of the lattice depths.

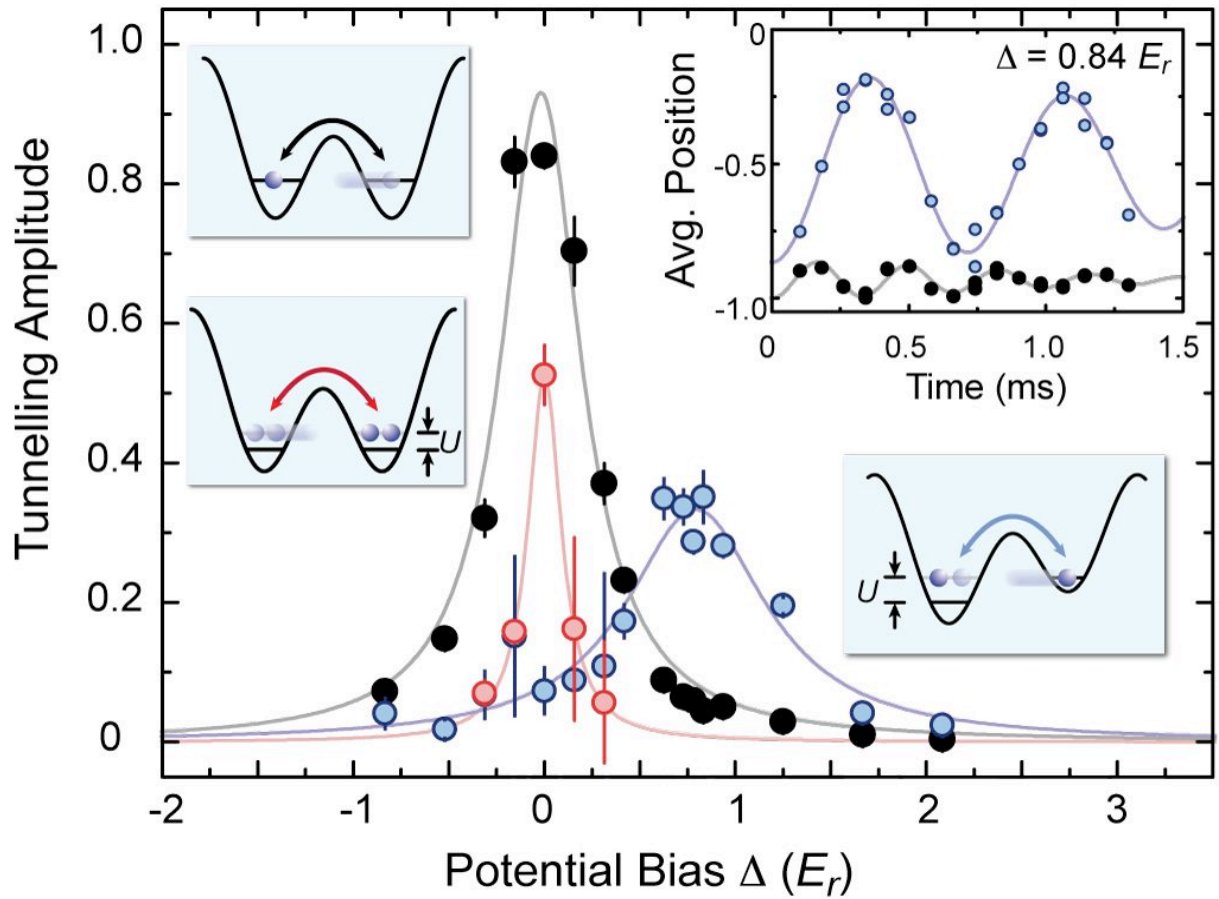

Fig. 4 Conditional tunnelling. The tunnelling amplitude vs. the potential bias is measured for the case of single atoms (black data points) and initially doubly occupied lattice sites (blue and red data points) for $V_{short}$=12 $E_r$, with $J/U$ ~ 0.2. The single atom tunnelling is only resonant in the unbiased case. For doubly occupied sites, two processes corresponding to first and second order tunnelling can be identified by their frequency. Each process shows a distinct resonance. The correlated pair tunnelling (red circles) is resonant for zero bias. Near a potential bias of $\Delta$=0.78(2) $E_r$ (centre of Lorentzian fitted to blue data points) a conditional tunnelling resonance occurs, where a single atom can tunnel only in the presence of a second atom. The dynamical evolution at this resonance can be seen in the inset, where the blue (black) data points denote the average atom position vs. time for the doubly (singly) occupied sites. Without a second atom the tunnelling is strongly suppressed. Amplitudes and error bars are derived from fit parameters and uncertainties of fitting

damped two-component sinusoidal functions to the position signal as shown in the inset.